\documentstyle[multicol,aps,epsfig,prb]{revtex}

\begin{document}

\draft

\title{Structure and relaxations in liquid and amorphous Selenium}

\author{D. Caprion and H.R. Schober}

\address{Forschungszentrum J\"ulich, IFF, D-52425 J\"ulich, Germany}

\date{\today}

\maketitle

\begin{abstract}
We report a  molecular dynamics simulation of selenium, described by a
three-body interaction. The temperatures $T_{\mathrm{g}}$ and
$T_{\mathrm{c}}$ and the structural properties are in agreement with
experiment. The mean nearest neighbor coordination number is  2.1. A small
pre-peak at about $1$~\AA$^{-1}$ can be explained in terms of void
correlations. In the intermediate self-scattering function, i.e. the density
fluctuation correlation, classical behavior, $\alpha$- and
$\beta$-regimes, is found. We also observe the plateau in the $\beta$-regime
below $T_{\mathrm{g}}$. In a second step, we investigated the heterogeneous
and/or homogeneous behavior of the relaxations. At both short and long times the
relaxations are homogeneous (or weakly heterogeneous). In the intermediate
time scale, lowering the temperature increases the heterogeneity. We
connect these different domains to the vibrational (ballistic), $\beta$-
and $\alpha$-regimes. We have also shown that the increase in heterogeneity
can be understood in terms of relaxations.
\end{abstract}


\begin{multicols}{2}

\section{Introduction}

Although glass is one of the most common materials, the glass transition is
still poorly understood. It is a continuous transition in which the viscosity
of the glass forming liquid increases from $10^{-3}$~Pa\,\,s in the liquid to
more than $10^{9}$~Pa\,\,s in the super-cooled state. It is, therefore, easy to
understand that very different time scales become important near the glass
transition, and different types of relaxations might be observed.

The Mode Coupling Theory (MCT) \cite{Goet} gives a microscopic
picture of this transition. This theory focuses on the density correlation
function $\Phi(q,t)$, the intermediate self-scattering function, and
proposes a mechanism of back-flow to explain the increase of the viscosity
and/or of the time-scales \cite{Barr}. One of the most striking results of
the MCT is the prediction of a critical temperature $T_c$ below which the
system becomes non-ergodic. In other words the system is trapped in a well
of the energy landscape. This feature is related to a non-zero value of
$\Phi(q,t)$ when $t \to \infty$. Above $T_{\mathrm{c}}$, the function
$\Phi(q,t)$ shows a short time relaxation, related to the vibrational (often
called ballistic) regime, and a long time one, also called
$\alpha$-relaxation. Below $T_{\mathrm{c}}$, a third regime appears, the
so-called $\beta$-regime, which is first seen as a shoulder and saturates at
a finite value below $T_{\mathrm{g}}$. 

This non-ergodicity has been seen in many experimental measurements
\cite{Meze,Alba,Rich,Sche} and computer simulations
\cite{Kob1,Teic,Yama,Affo} on different types of materials ranging from the
fragile polymeric glasses to strong glasses, such as SiO$_2$. In this
paper we want to go a step further. Using a model of selenium,
we check for
the presence of these two or three (depending on the temperature) relaxation
steps, and ask the following question. Does each atom have the same
probability to relax? If below $T_{\mathrm{c}}$ the system becomes
non-ergodic, and is trapped in a well of the energy landscape, are all
atoms still equivalent, or are some more (or less) mobile than 
others? We can reformulate this question and ask whether the relaxations are
homogeneous or heterogeneous. According to some authors \cite{col1,col2} the
relaxations should be more homogeneous, in particular in the $\alpha$-regime,
whereas others \cite{exp1,exp2,exp3,exp4,hard1,hard2,soft1,soft2,Kob2}
claim that the relaxations in amorphous or disordered materials are more
heterogeneous. The answer therefore does not seem to be obvious.

From the theoretical point of view, simple one-atomic systems such as soft or
hard spheres or Lennard-Jones systems would be optimal to study.
Unfortunately these simple systems crystallize rapidly at temperatures near
the glass transition temperature ($T_{\mathrm{g}}$) and, therefore, can
be utilized only for studies in the liquid, well above $T_{\mathrm{g}}$, or
deep in the glassy state, $T \ll T_{\mathrm{g}}$. One possibility to avoid
crystallization is to introduce special features in the inter-atomic
interaction potential which penalize ordering \cite{dzugutov:92,ruocco:00}.
The most common approach is to use binary mixtures of atoms. A different
approach is to simulate a real mono-atomic glass former, such as selenium,
which has been studied extensively in experiment, see the reviews
\cite{andonov:82,corb:82}. Se is covalently bound and prefers a coordination
number of two. This is reflected in the different crystal structures. The
most stable trigonal phase consists of parallel helical chains. Two
monoclinic forms are composed of rings of eight atoms. The polymorphs
are distinguished by the correlation between neighboring dihedral angles.
Depending on this correlation one has either a trans- (chains)
or a cis-configuration (rings). The energy difference between the cis- and
trans-configuration was estimated as only $0.03$~eV \cite{misawa:78}. This
low energy difference implies that in glasses both configurations should
coexist, which in turn strongly hinders crystallization.  From a
first-principles molecular dynamics simulation using 64 atoms Hohl and Jones
\cite{hohl:91} conclude that both amorphous and liquid selenium can be viewed
as consisting of branched chains which include rings of different length.
The small size prevented quantitative statistics of chain and ring lengths and
branching points. The fraction of atoms having twofold coordination varies
in the literature between 95\% and 71\% 
\cite{hohl:91,bichara:94,kirchhoff,shimizu:99}.  

To study dynamical properties, larger systems are needed and one has to resort
to effective inter-atomic interactions. This immediately leads to the
difficulty of simultaneously having to describe the covalent binding in the
chains and rings and the van-der-Waals like interaction between the rings, as
well as possible branching and bond breaking. One possibility is to disregard
the last two, and to use different interactions for atoms in the same chain 
and
in different chains, respectively. Similar to simulations of polymers one
then considers a glass or a liquid of chains of a predefined length. This
fragmented chain method has been employed both for electronic structure
calculations \cite{yamaguchi:99} and for classical molecular dynamics
simulations \cite{bermejo,balasub:92}. Another possibility is to use
a more simple description of the electronic properties, such as tight binding
models \cite{bichara:94}. 

We follow a different line. As done before by Stillinger {\it et al.} in
their study of liquid sulfur \cite{stillinger:86}, we use one effective
inter-atomic potential for both, the intra-chain and the inter-chain
interactions.

This paper is laid out as follows: in section II we report the details of
the simulations, and of the production of the liquid and glassy samples
used in the measurements of the different properties reported in this work.
Section III is devoted to the determination of the glass transition temperature
$T_{\mathrm g}$ and the critical temperature $T_{\mathrm c}$. Given these
temperatures, we report the evolution of the structure through the glass
transition in section IV.
In the next section we focus on the relaxations and the intermediate
self-scattering function. In section VI we present the tools, used to study
the  homogeneity or heterogeneity of these relaxations, report the
measurements and discuss them. Finally we conclude. 

\section{Simulations}

We performed classical molecular dynamics simulations on a system of $N=2000$
atoms interacting via a 3-body potential. This potential was built to
reproduce the properties of small clusters of selenium and of the
trigonal crystalline phase. Details of the potential are given in Ref.
\cite{Olig1}. The potential has previously been used to calculate the
vibrations \cite{OS:vib} and local relaxations in amorphous Se \cite{OS:rel}.
In these simulations it was shown that both the low frequency resonant
vibrations and the local relaxations are centered on groups of ten and more
atoms. This finding is in agreement with the interpretation of experiments
by the soft potential model \cite{BGGS:91}.
From a Monte Carlo study of liquid Se it was concluded that the model
provides a sound basis for the study of both the microscopic and the
electronic structure, despite some deficiency in the treatment of the
van-der-Waals interaction \cite{koslowski:99}.

During the simulations the pressure was fixed to zero Pa, i.e. we worked with
equilibrium structures. In order to keep the pressure constant we used a
Parrinello-Rahman algorithm \cite{Parr1,Parr2}. The temperature was 
kept constant by rescaling the velocities at each integration step. 
We checked that the way we control the pressure and temperature 
influenced neither the dynamics of the system nor the results of our simulations.

The equations of motion were integrated using the velocity Verlet algorithm
\cite{Swop}. Taking care of the stability of the algorithm, we chose the time
steps equal to $1$~fs in the liquid, $2$~fs in the glassy phase and $4$~fs
for the lowest temperature ($6$~K).
 
To improve the statistics we used four independent starting 
configurations to obtain the different samples used in the measurement.
These samples were produced as follows: first we equilibrated a liquid at
$550$~K (above the melting point), and cooled it to $290$~K with a
quench rate of $10^{13}$~K/s. At this temperature we let the systems
equilibrate during $8$~ns and then quenched them to $0$~K with the
same quench rate. During both quenches we saved configurations at
several temperatures and again equilibrated them before using them as 
input of the measurements. The equilibration times were $8$~ns above 
$290$~K, $16$~ns between $290$~K and $6$~K, and $32$~ns at $6$~K. After 
the equilibration period several relevant physical quantities were computed.

\section{T\lowercase{$_{\mathrm g}$} and T\lowercase{$_{\mathrm c}$}
determination}

To obtain the relevant temperature scale, we first determined the glass
transition temperature $T_{\mathrm g}$. For this we followed the evolution of
the volume during the quench process. As the coefficient of volume expansion
is higher in the liquid than in the solid phase, one observes a change of
slope of the volume curve  when the system is quenched through the glass
transition. 

\begin{figure}
\centerline{
  \hbox{\epsfig{figure=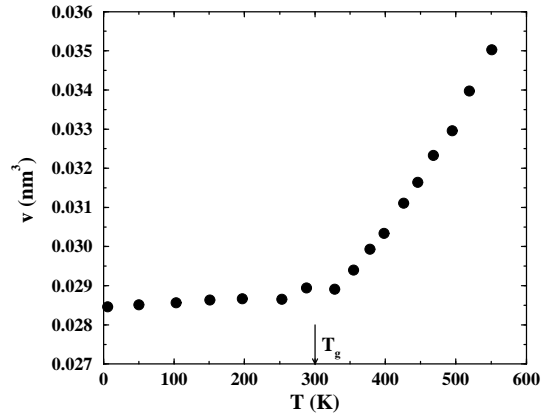,width=7cm}}
}
\caption{Evolution of the atomic volume of liquid and amorphous 
Se atoms during the 
quench. The change of slope
between high and low temperatures determines the glass transition 
temperature $T_g$.}
\end{figure}

From Fig.~1 the glass transition temperature is estimated as
$T_{\mathrm g} \approx 300$~K.
Experimentally $T_{\mathrm g}$ is about $305$~K \cite{Buch}. 
The good agreement between the simulated and experimental values of $T_g$
might seem surprising regarding the high quench rate and the usual
discrepancies observed in simulation. However, one should notice that 
due to the aging over several ns the effective quench rate is lower,
$\approx 10^{10}$~K/s.

Another characteristic temperature is the critical temperature
$T_{\mathrm c}$ given by the Mode Coupling Theory. This temperature can be
obtained from the diffusion constant $D$ which, according to the MCT, follows
a power law $(T-T_{\mathrm{c}})^\gamma$ \cite{Goet}. The diffusion constant
is related to the atomic mean square displacement by
\begin{equation}
D=\displaystyle \lim_{t \to \infty} \frac{\left< |{\mathbf r}(t+t_0)-{\bf r}
(t_0)|^2 \right>_{t_0}}{6t}.
\end{equation}
The diffusion constants obtained, Fig.~2, are in reasonable agreement
with experiments on liquid Se \cite{Phil,axmann:70}. At the higher temperatures
$D$ can be fitted by an
Arrhenius law with an activation energy of $0.3$~eV, in agreement with
results from first-principle molecular dynamics simulations \cite{kirchhoff}.
In the inset of Fig.~2 we also show, by a dashed line, the fit by the MCT power
law. Due to the correlation between $T_{\mathrm{c}}$ and $\gamma$ such a fit can
only approximately determine these values. 

\begin{figure}
\centerline{
  \hbox{\epsfig{figure=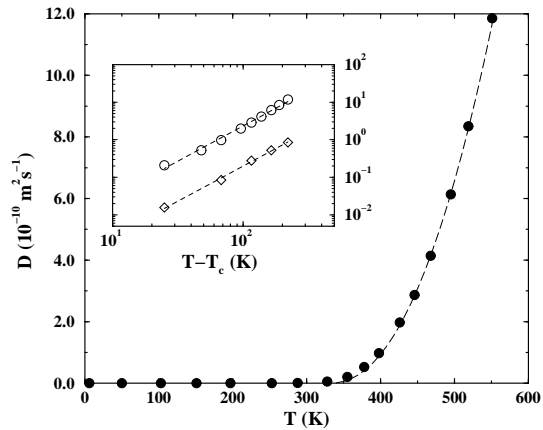,width=7cm}}
}
\caption{Diffusion constant of liquid and amorphous Se as function of 
temperature. The full circles show the values obtained from the MD 
simulations. The dashed line shows a fit with the power law
$D\propto(T-T_c)^\gamma$. The inset shows the diffusion constant D ($\circ$)
and the decay time of the $\alpha$-relaxation $\tau^{-1}$ ($\diamond$) versus
$(T-T_c)$ in a log-log representation, the y-axis is in units of
$10^{-10}$~m$^2$s$^{-1}$ for $D$, and ps$^{-1}$ for $\tau^{-1}$. }
\end{figure}

Fixing $T_{\mathrm{c}}$ about 10\% above $T_{\mathrm g}$, i.e.
$T_{\mathrm{c}} = 330$~K we get $\gamma = 1.88$. From the same fit to the
decay time of the $\alpha$-process, see section V, we obtain for
$T_{\mathrm{c}} = 330$~K a value $\gamma = 1.86$, 
which is in excellent agreement.

\section{Structural properties}

Having obtained the relevant temperature scales we now turn to
structural properties. First we compute the pair correlation function (PCF)
at temperatures ranging from the liquid down into the glass. The
PCF is defined by:
\begin{equation}
g(r)={\displaystyle\frac{V}{4\pi r^2 N^2}}
\left<\sum_{i}\sum_{j\neq i}\delta(r-r_{ij})\right>
\label{eq_gr}
\end{equation}
where $\left< ... \right>$ denotes the average over configurations.

\begin{figure}
\centerline{
  \hbox{\epsfig{figure=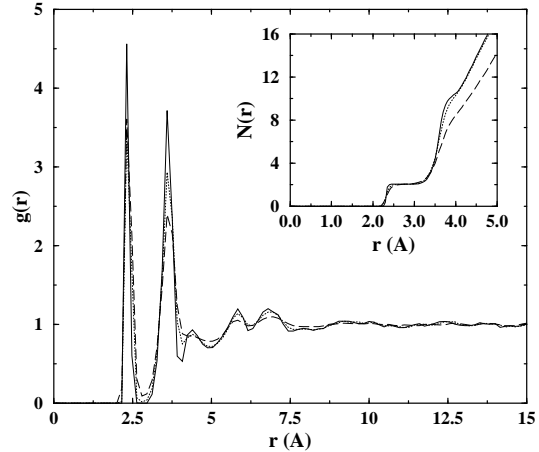,width=7cm}}
}
\caption{Pair correlation function of Se at three different temperatures: 
$6$~K (solid line), 
$290$~K (dotted line), $550$~K (long dashed line). The inset shows the mean 
coordination numbers for
the same temperatures.}
\end{figure}

Fig.~3 shows the PCF for three different temperatures, $550$~K in the
liquid phase, $290$~K just below $T_{\mathrm g}$ and $6$~K 
deep in the glassy phase. The inset shows the mean coordination
number. The positions of the peaks do not change strongly upon cooling, but
broaden markedly. As usual oscillations at large distances are more
strongly damped at high temperatures. The spatial
correlations at large distances weaken with increasing temperature. In all
cases the correlations are weak for distances greater than $10$~\AA.
The minimum near $4$~\AA\ for low temperatures signals that
the second neighbor shell becomes more pronounced. The mean coordination
number (Fig.~3 inset) indicates the same behavior. The mean nearest
neighbor coordination is about $2.1$ at all temperatures, similar to the
experimental value\cite{Edel}. This value of around 2 is the signature of
the chains and rings forming the amorphous selenium structure. The small
excess of $0.1$ compared to the ideal value of 2 indicates a prevalence of
over-coordinated atoms (branching) over under-coordinated ones (chain ends).
At the lowest temperature ($6$~K) we also observe a small plateau in
the coordination number near $4$~\AA. The change of neighbor numbers with
temperature for larger distances reflects the lower density at high
temperatures.
This indicates that with increasing temperature the chain
structure remains, but the distance between chains increases.

From the PCF the structure factor $S(q)$ can be computed by a spatial
Fourier transform:
\begin{equation}
S(q)= 1 + \displaystyle \frac{V}{N} \int_0^\infty 4\pi r^2 (g(r)-1) 
\frac{\sin(qr)}{qr}dr
\end{equation}

Fig.~4 shows $S(q)$ for the three temperatures used in Fig.~3. As in the PCF,
the peaks do not shift strongly with temperature, they merely become more
damped with
increasing temperature. The positions of the peaks agree with
experiments \cite{Edel,Tamu} and previous simulations on
Se \cite{kirchhoff,Alma}. In addition we see a small pre-peak near
$q=1$~\AA$^{-1}$. Experimentally the pre-peak in amorphous selenium merges
with the first diffraction peak at about $2$~\AA$^{-1}$ and is only seen
as a shoulder.

\begin{figure}
\centerline{
  \hbox{\epsfig{figure=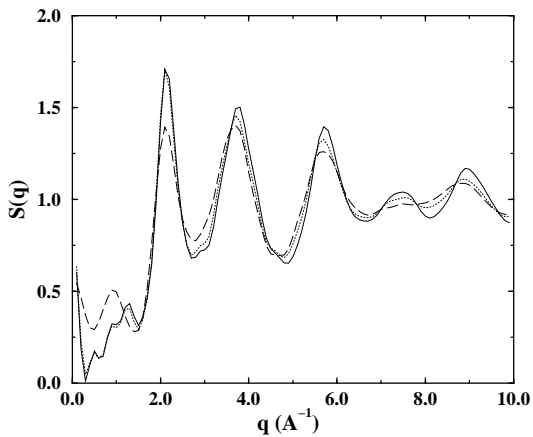,width=7cm}}
}
\caption{Structure factor of Se at the same temperatures as Fig. 3: 
$6$~K (solid line), $290$~K
(dotted line), $550$~K (long dashed line).}
\end{figure}

To study this pre-peak, we quenched two more sets of 10 independent liquids,
each to zero K applying two different pressures: zero pressure and  $10$~GPa.
Finally 
we minimized the potential energy for both sets, and released the pressure 
for the second set. This gave us at $T=0$~K two sets of samples with different
densities $\rho=4400$~kg/m$^3$ and $\rho=5090$~kg/m$^3$, both with equilibrium
structures. The average potential energy per atom of the low density samples is
only $3.5$~meV less than the one at the high density. This low value might
indicate that at high temperature voids are present in thermodynamic
equilibrium.
For both sets of equilibrium structures we computed the structure factors by
\begin{equation}
S(q)=\left<\sum_{i,j}\exp(i{\bf q}({\bf r}_j(t)-{\bf r}_i(t)))
\right>_{t,|{\bf q}|=q},
\end{equation}
where ${\bf q}$ are the $q$-vectors compatible with the simulation box. The 
minimal $q$-values, given by the periodic boundary conditions are $q=0.17$
\AA$^{-1}$ and $q=0.16$~\AA$^{-1}$ for the high and the low density samples,
respectively.

\begin{figure}
\centerline{
  \hbox{\epsfig{figure=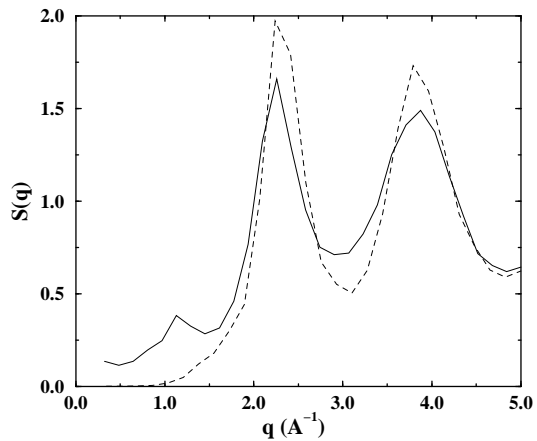,width=7cm}}
}
\caption{Structure factor of Se at $T = 0$~K for two different densities: 
$\rho=4400$~kg/cm$^3$
(solid line) and $\rho=5090$~kg/cm$^3$ (dashed line).}
\end{figure}

Whereas the low density samples show a small pre-peak near $1$\AA$^{-1}$, it is
absent in those of
high density (Fig.~5). Checking the mean coordination number at
the two  densities (Fig.~6), one sees that the number of first neighbors
changes very little with density: there are chains and rings at both densities.
The number of second neighbors, however, is reduced for the lower density. 

\begin{figure}
\centerline{
  \hbox{
    \epsfig{figure=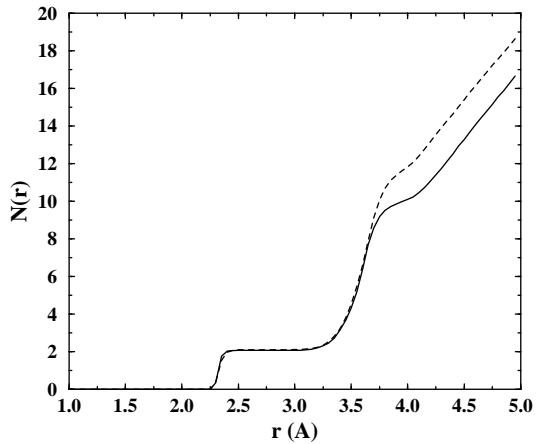,width=7cm}
  }
}
\caption{Mean coordination number of Se at $T = 0$~K at two different 
densities: $\rho=4400$~kg/cm$^3$
(solid line) and $\rho=5090$~kg/cm$^3$ (dashed line).}
\end{figure}

This
is the same effect as previously noted for the temperature dependence. When the
density is low, i.e. when the volume is high the system is constituted of chains
and rings which are further apart from each other. In other words we interpret
the pre-peak as the signature of correlations between holes in the structure.
A similar effect was seen in simulations of SiO$_2$ \cite{elliott:95}.
As a consequence of the too high density of the crystalline structure 
\cite{Olig1}, the density of our amorphous selenium is too high in comparison
with the experimental value.

\section{Intermediate self-scattering function}

We now focus on the local 
relaxations. First, we compute the intermediate self-scattering function
(ISSF), the correlation function of the density fluctuations,
\begin{equation}
\Phi(q,t)=\displaystyle\left<\delta\rho_{-q}(t+t_0)\delta\rho_{q}(t_0)
\right>_{t_0}.
\end{equation}
This can be rewritten as the spatial Fourier transform of the van
Hove self correlation function $G_{\mathrm{s}}(r,t)$
\begin{equation}
\Phi(q,t)= \displaystyle \int_{0}^{\infty} G_{\mathrm{s}}(r,t)
\frac{\sin(qr)}{qr} dr
\end{equation}
where $G_{\mathrm{s}}(r,t)$ is given by \cite{VanH}:
\begin{equation}
G_{\mathrm{s}}(r,t) =\displaystyle\left<\delta (r-|{\bf r}_i(t+t_0)-
{\bf r}_i(t_0)|) \right>_{i,t_0}.
\end{equation}

The ISSF of Se is not easily accessible in experiment. It corresponds
to the time Fourier transform of the incoherent dynamic structure factor,
but neutron scattering by Se is mainly coherent. Nevertheless this
quantity is accessible to simulation.

\begin{figure}
\centerline{
  \hbox{\epsfig{figure=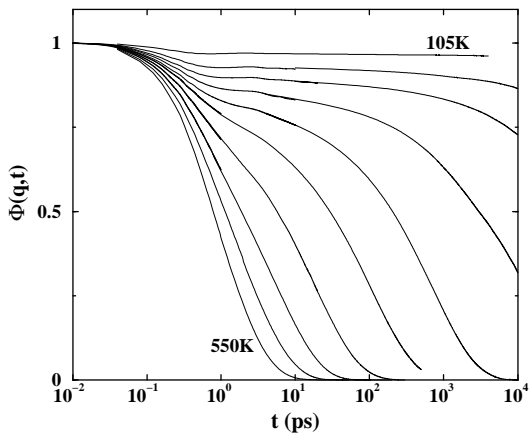,width=7cm}}
}
\caption{Intermediate self-scattering function (ISSF) of Se versus time for 
different temperatures
above and below $T_g$. From top to bottom: $105K$, $200K$, $255K$, 
$290$~K,
$330$~K, $355$~K, $400$~K, $445$~K, $495$~K and $550$~K.}
\end{figure}

In Fig.~7 we report the ISSF at $q=2.1$~\AA$^{-1}$ 
corresponding to the first diffraction peak of the $S(q)$ in Fig.~4. 
A fast decrease of $\Phi(q,t)$ on the time-scale of a picosecond, 
is observed at all temperatures. Decreasing the temperature a shoulder
appears for  intermediate time scales. As the system approaches
the glass transition temperature, $\Phi(q,t)$ starts to saturate and finally
shows a plateau for intermediate and long times. As customary this curve
is described as follows: First there is the ballistic or vibrational regime 
(corresponding to the fast decreases at low time). Then, for
$T > T_{\mathrm{c}}$, $\Phi(q,t)$ goes to zero (the so-called
$\alpha$-regime). The shoulder or plateau at lower $T$ corresponds to
the so-called $\beta$-regime. This plateau indicates that the system is trapped
in a limited area of phase-space. 

According to the MCT, the $\alpha$-regime above $T_{\mathrm{c}}$ is driven by
a master curve which is obtained by rescaling the time by a characteristic
decay time, $\tau$ defined as the time when the ISSF has drops to $1/e$, 
$\Phi(q,\tau) = 1/e$. Above $T_{\mathrm{c}}$, similarly to
the diffusion constant $D$ these values $\tau(T)$ should obey a power law
$\tau(T) = (T-T_{\mathrm c})^{-\gamma}$. Fixing $T_{\mathrm{c}}=330$~K we get a
good fit with $\gamma = 1.86$, see inset in Fig.~2. Using this $\tau$ the master
curve can be written in scaled form as
\cite{Barr}:
\begin{equation}
\Phi(q,t/\tau)  = \Phi_0 \exp(-(t/\tau)^\beta).
\end{equation} 
Fig.~8 presents the curves for temperatures above $T_{\mathrm g}$
rescaled by  their respective $\tau$. We get a value $\beta = 0.53$ for
temperatures ranging from $T=330$~K to $T=430$~K. We do not intend to give an
elaborate test of the MCT but show the rescaled curves merely to indicate
that the rescaling procedure holds approximately.

\begin{figure}
\centerline{
  \hbox{\epsfig{figure=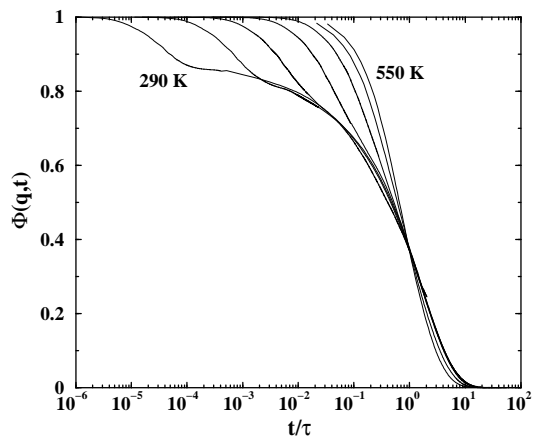,width=7cm}}
}
\caption{Intermediate self-scattering function of Se versus the 
dimensionless variable 
$t/\tau$, where $\tau$ is defined by $\Phi(q,\tau)=1/e$, see inset of Fig. 2.
Temperatures from left to right: $290$~K, $330$~K, $355$~K, $400$~K, 
$445$~K, $495$~K and $550$~K.}
\end{figure}

The most striking effect is the plateau corresponding to the $\beta$-regime. 
It indicates that the system falls out of equilibrium, 
and that atoms are trapped in
a well of the energy landscape. This poses the question 
of whether all the atoms
are affected equally or not.

The same question can be asked for the $\alpha$-regime which can be represented
by a stretched exponential decay law.
Such a law can either mean that a stretched exponential
decay law governs the whole system, or it originates from an average of normal
exponential laws with different time constants. The first picture is called
homogeneous scenario (the system is everywhere the same) and the second one
heterogeneous.
 
\section{Heterogeneity or homogeneity ?}

To answer this question, we use again the van Hove correlation function (VHF)
which represents the probability for an atom to move a
distance $r$ during a time $t$. If all the atoms have the same mobility the
VHF is a Gaussian multiplied by the geometrical factor $4\pi r^2$. In the
following this factor is always thought to be included when we speak of
Gaussianity.  In the opposite case if some atoms are trapped or some are more
mobile than the majority the VHF will, in general, be non-Gaussian.
As example we show in Fig.~9 the VHF for two different temperatures for the
same length range but for two different times. Obviously at high temperatures
the system has a higher diffusion constant  and the atoms will move faster
over a given distance. More striking is that at high
temperatures (in the liquid) the VHF has Gaussian shape, whereas at low
temperatures (in the glass) the VHF has an extended tail and cannot be
described by a Gaussian. Some atoms move over much larger distances than the
average atom, i.e. they have a much higher mobility.

\begin{figure}
\centerline{
  \hbox{\epsfig{figure=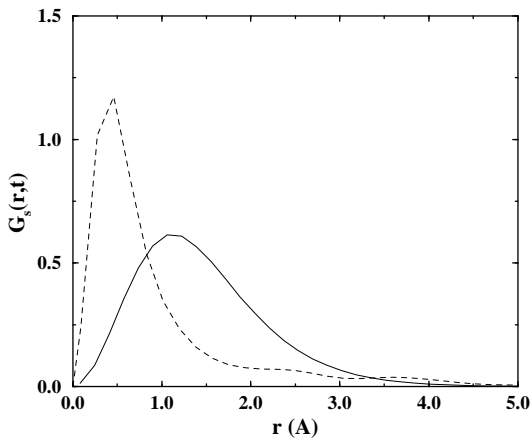,width=7cm}}
}
\caption{Van Hove correlation function of Se at two different 
temperatures (solid line: 
$550$~K; dashed line $255$~K) and times, $t=2.1$~ps for $550$~K and
$t =36$~ns for $255$~K, respectively. The different times reflect the 
higher mobility (or diffusion) in the liquid.}
\end{figure}

In order to quantify these findings, and in accordance with previous work, we
introduce the non-Gaussianity parameter (NGP) $\alpha_2$ \cite{Rahm1}
\begin{equation}
\alpha_2(t)=\displaystyle\frac{3\mu_4}{5{\mu_2}^2}-1,
\end{equation}
where $\mu_2$ and $\mu_4$ are the second and fourth moments of the VHF, 
$\mu_2 =\left<r^2(t)\right>$ and $\mu_4 = \left<r^4(t)\right>$, respectively.
The NGP is identical  to zero for a Gaussian VHF.  

Fig.~10 shows, in a log-linear representation, the $\alpha_2$ at different
temperatures from the liquid to the glass for a time span covering 6 decades. 
At very short times the NGP is nearly zero at all temperatures. The limiting
behavior for large times is more difficult to observe, especially at low
temperatures. Nevertheless the NGP clearly tends to zero. Furthermore, all the
curves, below and above the glass transition, scale to the same curve
in the short time range as already shown by Kob {\it et al.} in their study
of a binary super-cooled Lennard-Jones liquid \cite{Kob2}. Our
work shows that this property persists in the glassy phase and for
a completely different structure. 

\begin{figure}
\centerline{
  \hbox{\epsfig{figure=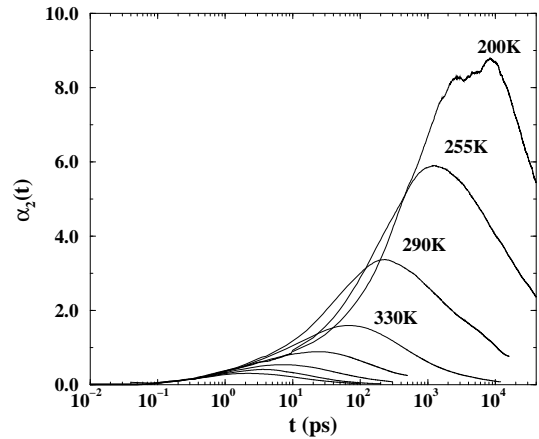,width=7cm}}
}
\caption{Log-linear plot of the non-Gaussianity parameter, $\alpha_2$, 
of Se versus time for several 
temperatures. From top to 
bottom : $200$~K, $255$~K, $290$~K, $330$~K, $355$~K, $400$~K, $445$~K 
and $495$~K.}
\end{figure}

In the intermediate time range the NGP has a maximum indicating heterogeneity.
At high temperatures, in the liquid above $T_{\mathrm c}$, this maximum is
small and located around $10$~ps. Upon cooling, it slowly moves to higher times.
For the temperatures below $T_{\mathrm c}$, the behavior is different. The value
of the maximum is larger than 2 and it grows by a factor 2 upon cooling by
$50$~K. Similarly the position of the maximum shifts by about an order of
magnitude for each $50$~K. These two observations suggest that as the system is
cooled especially below $T_{\mathrm g}$, the non-Gaussianity becomes more
and more pronounced at intermediate time scales.

\begin{figure}
\centerline{
  \hbox{\epsfig{figure=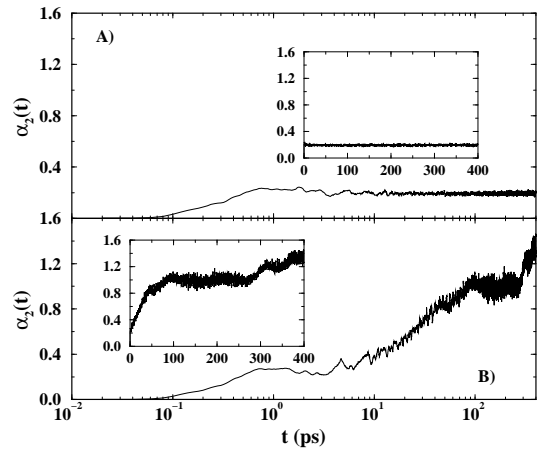,width=7cm}}
}
\caption{Log-linear plot of the non-Gaussianity parameter of Se versus time 
for two samples 
(A and B) at low
temperature: $6$~K. The insets show  the same quantities in a 
linear-linear plot.}
\end{figure}

We now focus on the short time behavior at very low temperatures. In
Fig.~11, we present the evolution of the NGP for two
different samples A and B (out of our 4 different samples) at a very low
temperature, $T=6$~K. The inset
gives the same curves in linear-linear representation to show them
clearly at intermediate times. The curves coincide
during the first ps in the vibrational regime. For the
larger, intermediate time scale the NGP of sample A
(Fig.~11.a) oscillates around a value of 0.2, but the one of sample B
(Fig.~11.b) rises. The two other samples behave similarly to
sample A. What is the reason of this difference in the
non-Gaussian behavior of these two kinds of samples? The evolution of both 
total energy and volume were equivalent. 

\begin{figure}
\centerline{
  \hbox{\epsfig{figure=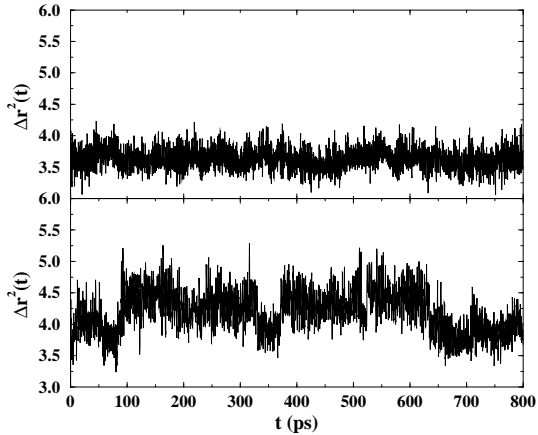,width=7cm}}
}
\caption{Mean square displacement for samples A (top) and B (bottom) during 
the simulation.}
\end{figure}

The mean square displacements, however,
evolve differently (fig. 12). In sample A (Fig.~12.a) it oscillates around a 
mean value during the entire simulation run, whereas it shows steps for sample B
(Fig.~12.b). Thus, while sample A stays in one well of the energy landscape,
sample B moves from one well into another. We can identify at least two
different wells for sample B. We conclude that the value $\alpha_2 \approx 0.2$
of the NGP corresponds to the vibrations in the liquid and amorphous selenium.
Relaxations from one minimum of the energy landscape to another lead to an
increase in the NGP. It has been shown previously that that these local
relaxations are collective jumps of 10 and more atoms.\cite{OS:rel}
The jump length of a single atom is much less than the nearest neighbor 
distance. The different behavior of the samples, shown in Fig.~12
reflects the low
probability for relaxations at low temperature.  It
is not restricted to Se but is typical for glasses. The same has been
reported also from a simulation of simple soft sphere glass \cite{OS:99} and
is observed experimentally in the telegraph noise of the electric resistivity
of point contacts \cite{kozub:98}. 
  
\section{conclusion}

In this article we have presented results of a molecular dynamics simulation
on the structure and relaxations of liquid and amorphous Se. We determined
the glass transition and critical temperatures, the pair correlation
function and the structure factor. From the pair correlation function, and in
agreement with experiments, we concluded that both liquid and amorphous
selenium are constituted of chains and rings with a mean coordination number 
of 2.1, slightly above the ideal value 2. Rings and chains are
interconnected. The structure factor shows a small pre-peak around
$1$~\AA$^{-1}$ which in experiment only shows as a shoulder of the main peak.
This pre-peak is explained in terms  of a correlation of
voids between the selenium chains. To prove this assumption we
computed the structure factors of two sets of samples with two different
densities. At the higher density no pre-peak is observed. 

The van Hove correlation function was calculated and utilized to compute
the intermediate scattering function and the
non-Gaussianity parameter. For the intermediate
self-scattering function, the time correlation of the density fluctuation,
we find the classical behavior:
at short times a rapid decrease corresponding to the
ballistic (or vibrational) regime, and at long times a slow decay
corresponding to the $\alpha$-regime.  When the system reaches $T_{\mathrm c}$
a shoulder and below $T_{\mathrm g}$ a plateau evolves between these two 
regimes. This corresponds to
the $\beta$-regime, and to a memory effect of the correlation function, in
other words the system falls out of equilibrium. 

The non-Gaussianity parameter shows that,  both at short and
long times, the relaxations are homogeneous or only weakly inhomogeneous 
and all the atoms are largely equivalent. For the intermediate time
range (corresponding to the $\beta$-regime) NGP depends on the
temperature. The lower the temperature the higher the NGP, i.e. the higher 
the heterogeneity. At low temperatures  
the increase of non-Gaussianity is due to relaxations. We conclude the
following scenario for the heterogeneity: at all temperatures, both above and
below $T_g$, there is a small increase of heterogeneity 
($\alpha_2 = 0.2$) due to
vibrations at short time, at intermediate times a pronounced
increase, due to the relaxations especially at temperatures below $T_g$, 
and finally a
decrease, due to long
range diffusion (flow motion). These different regimes correspond to the
different regimes observed in the 
intermediate self-scattering function $\Phi(q,t)$. 
This scenario seems to be common to different materials.

\section{Acknowledgment}

We would like to thank C. Oligschleger for her help at the beginning of 
this work.
We also thank M. Kluge, J. Matsui and U. Buchenau for fruitful
and exciting discussions.
One of us (D. C.) is grateful to the A. von Humboldt foundation for 
financial support.

\end{multicols}

\end{document}